\documentclass[%
 reprint,
superscriptaddress,
 amsmath,amssymb,
 aps,
]{revtex4-1}

\usepackage{graphicx}% Include figure files
\usepackage{dcolumn}% Align table columns on decimal point
\usepackage{bm}% bold math
\usepackage[english]{babel}
\usepackage{color}
\usepackage{xr}
\usepackage{subfigure}

\begin{document}

%\preprint{APS/123-QED}

\title{Updated Constraints on Asteroid-Mass Primordial Black Holes as Dark Matter}% Force line breaks with \\

\author{Nolan Smyth}
\email{nwsmyth@ucsc.edu}
\author{Stefano Profumo}%
 \affiliation{Department of Physics, University of California, Santa Cruz, CA 95064, USA}
 \affiliation{Santa Cruz Institute for Particle Physics, Santa Cruz, CA 95064, USA}
 \author{Samuel English}%
  \affiliation{Department of Physics, University of California, Santa Cruz, CA 95064, USA}
 \author{Tesla Jeltema}%
 \affiliation{Department of Physics, University of California, Santa Cruz, CA 95064, USA}
 \affiliation{Santa Cruz Institute for Particle Physics, Santa Cruz, CA 95064, USA}
\author{Kevin McKinnon}
\author{Puragra Guhathakurta}
\affiliation{Department of Astronomy and Astrophysics, University of California, Santa Cruz, CA 95064, USA}

\date{\today}% It is always \today, today,

\begin{abstract}
Microlensing of stars places significant constraints on sub-planetary-mass compact objects, including primordial black holes, as dark matter candidates. As the lens' Einstein radius in the source plane becomes comparable to the size of the light source, however, source amplification is strongly suppressed, making it challenging to constrain lenses with a mass at or below $10^{-10}$ solar masses, i.e. asteroid-mass objects. Current constraints, using Subaru HSC observations of M31, assume a fixed source size of one solar radius. However, the actual stars in M31 bright enough to be used for microlensing are typically much larger. We correct the HSC constraints by constructing a source size distribution based on the M31 PHAT survey and on a synthetic stellar catalogue, and by correspondingly weighting the finite-size source effects. We find that the actual HSC constraints are weaker by up to almost three orders of magnitude in some cases, broadening the range of masses for which primordial black holes can be the totality of the cosmological dark matter by almost one order of magnitude.
\end{abstract}

\maketitle

%\tableofcontents

\section{Introduction}

The microscopic nature of the dark matter (DM) permeating and shaping the observed universe remains mysterious. The persistent lack of a direct signal from weak-scale DM particle candidates
\cite{arcadi_waning_2017} has spurred growing interest in other possibilities. Primordial black holes (PBH), formed in the early universe as a result of large primordial density perturbations, are compelling DM candidates if massive enough to survive Hawking evaporation over the age of the universe ($m_{\rm PBH}\gg 10^{-17}\ M_\odot$) \cite{carr_new_2010, boudaud_voyager_2019}. A variety of constraints, including, but not limited to, the effects of partial evaporation at low masses and microlensing at larger masses rule out PBHs contributing 100\% of the DM over most of the possible parameter space, roughly $5\times 10^{-17}<m_{\rm PBH}/M_\odot<10$. At present the only remaining window is towards the low-mass end and perhaps at the solar-mass end \cite{bird_did_2016}, although the latter might be constrained by CMB distortion caused by matter accretion onto the PBHs \cite{poulin_cmb_2017}, X-ray data which places limits on the photon flux from PBHs interacting with the interstellar medium \cite{inoue_new_2017}, and other microlensing and dynamical heating constraints (see e.g. \cite{green_microlensing_2016} and the references therein).

Here, we intend to correct and update constraints in the most plausible region where PBH could be the DM, i.e. the asteroid-mass range $5\times 10^{-15}<m_{\rm PBH}/M_\odot<10^{-10}$. Over this mass range, the role of finite-size source effects is critical: as the Einstein radius of the lensing object in the source plane becomes comparable to, or smaller than the source size, the source amplification from lensing is strongly suppressed \cite{ulmer_femtolensing:_1995}. 
Ref.~\cite{katz_femtolensing_2018} showed that GRB femto-lensing constraints, after correction for finite-size effects, do not presently constrain PBH as DM candidates; Ref.~\cite{niikura_microlensing_2019} (see also \cite{sugiyama_revisiting_2019}) realized their original constraints from optical observations of M31 had been vastly over-estimated because finite-source-size effects had originally not been accounted for. Ref.~\cite{montero-camacho_revisiting_2019} re-assessed the effects, pointing out that even in the corrected version the assumed source size might have been underestimated, leading to incorrect, very optimistic constraints.

\section{Methodology}

\subsection{Microlensing Formalism}

Gravitational lensing of astrophysical objects is a powerful tool for observing dark, massive objects \cite{paczynski_gravitational_1986}. For low-mass lenses, such as asteroid-mass PBHs, the images formed by gravitational lensing cannot be fully resolved. The result is that the source is magnified by a factor
\begin{equation}
A = \frac{\phi}{\phi_0},
\end{equation}
where $\phi_0$ is the flux in the absence of lensing. The relevant scale for lensing by a PBH is the Einstein radius, $R_E$, which is defined as
\begin{equation}
    R_E = \sqrt{\frac{4GM_{PBH}d_L(1- d_L/d_S)}{c^2}},
\end{equation}
where $d_S$ and $d_L$ are the distances between the observer and the source, and the observer and the lens respectively. The angular size of the Einstein radius is given by $\theta_E \equiv \frac{R_E}{d_L}$ and represents the angle between the source and its image as measured by an observer when the lens is directly between the observer and source.

If we ignore the effects of wave optics (geometric optics approximation)  the magnification for a point source can be shown to be \cite{nakamura_wave_1999} 
\begin{equation}
    A_{\rm geo} = \frac{u^2 + 2}{u\sqrt{u^2 + 4}},
    \label{Ageo}
\end{equation}
where  $\boldsymbol{\theta}$ is defined to be the angle of the source such that $ u = \boldsymbol{\theta} / \theta_E$ is now the dimensionless impact parameter. More generally, we can define the following dimensionless quantities
\begin{equation}
    \textbf{x} = \boldsymbol{\theta}_L / \theta_E, \hspace{3mm}
    \textbf{y} = \boldsymbol{\theta} / \theta_E,\hspace{3mm}
    w = \frac{d_L d_S}{d_{LS}}\theta^2_E\omega,
    \label{Amplification}
\end{equation}
where $\theta_L$ is the angular size of the lens, $d_{LS}$ is the distance between source and lens, $\omega$ is the frequency of the light being lensed, and $\theta_E$ is chosen to be the characteristic angular scale. The general amplification $A$ for the case of a spherically symmetric lens reads \cite{nakamura_wave_1999}
\begin{equation}
    A = -iwe^{\frac{iwy^2}{2}} \int_0^{\infty} x J_0(wxy)e^{iw\big(\frac{x^2}{2} - \psi(x)\big)} dx,
\end{equation}
where $J_0$ is the zeroth order Bessel function. The Schwarzschild radius of a PBH is very small compared to the distances between M31, the PBH, and the HSC. Thus, the point mass lens approximation is valid and the amplification becomes
\begin{equation}
    A \simeq \frac{\pi w}{1 - e^{-\pi w}}\Big| {}_1F_1\Big(\frac{1}{2}i w,1;\frac{1}{2}i w y^2\Big)\Big|^2,
    \label{hyper}
\end{equation}
where ${}_1F_1$ is the confluent hypergeometric function \cite{nakamura_wave_1999, takahashi_wave_2003}.

In the short wavelength limit ($w \gg$ 1), the integrand of Eq. (\ref{hyper}) oscillates rapidly and the biggest contribution to the integral comes from stationary points corresponding to geometric images of Eq. (\ref{Ageo}). When $w \geq 1$, the geometric optics approximation breaks down and the wave diffraction effect becomes significant.

\begin{figure}[!h]
    \begin{center}
        \includegraphics[scale=0.28]{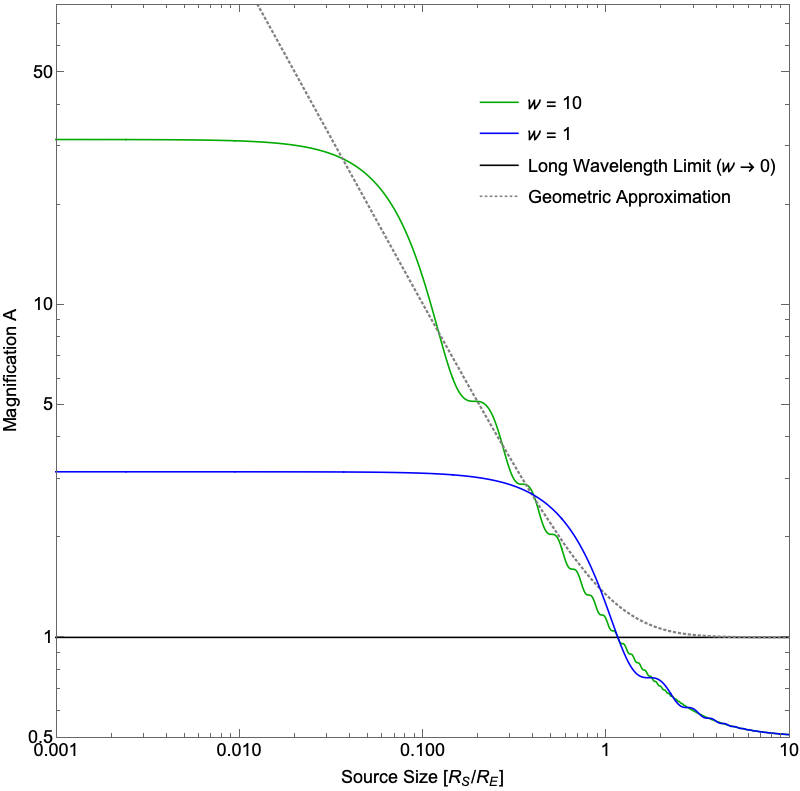}
        \caption{The magnification for different values of the dimensionless frequency $w$. In the geometric approximation, shown in dotted gray, the wave effects of light are ignored and the magnification is independent of the source size. In the long wavelength limit, shown in solid gray, the light effectively ignores the lens and no magnification occurs. For $w \geq 1$, the magnification increases as source size decreases, but reaches a maximum at $\pi w$.}
        \label{FiniteWaveOmega}
        \end{center}
\end{figure}

Previous microlensing constraints were obtained using observations from the Subaru telescope with the Hyper Suprime Cam (HSC) in the r-band filter \cite{niikura_microlensing_2019}. The wavelength of light detectable in this band is comparable to the Schwarzschild radius of a PBH with $M_{PBH} \leq 10^{-11} M_{\odot}$ . As a result, the geometric lensing approximation is not appropriate and diffraction effects must be taken into account. This renders the very low mass range untestable by the HSC. The HSC is sensitive to a microlensing event when the magnification is $A \geq 1.34$. At small values of the dimensionless frequency $w$ (the long wavelength limit), there can be no detectable lensing events because the magnification will always be lower than this threshold as shown in Fig.~\ref{FiniteWaveOmega}.

We note that the previous microlensing constraints show no appreciable difference between wave and geometric approaches when the finite size effect is taken into account \cite{sugiyama_revisiting_2019, niikura_microlensing_2019}. This is because during a real observation, the diffraction term in Eq. (\ref{Amplification}) averages out. In this case, the diffraction effects provide only small corrections to the threshold impact parameter found by only considering finite size effects. Therefore, in this paper we limit our discussion to the dominant effects of finite source size.

\subsection{Finite Size Effects}

The microlensing magnification generically depends on the size of the source \cite{matsunaga_finite_2006}. The source size, along with the magnitude of the impact parameter, determines whether the peak magnification for an extended source is enhanced or diminished \cite{witt_can_1994}. When the impact parameter becomes small ($u \ll r_{source}$), the peak magnification for a point source diverges, while the peak magnification for an extended source remains finite. This means that for extended sources, the peak magnification is significantly lower than in the point source case. We recalculate the magnification, taking into account the finite size effects following \cite{witt_can_1994} and \cite{sugiyama_revisiting_2019}. 

To determine the magnification of a finite-size source, we use the size of the source in the plane of the lensing PBH. It is convenient to define the parameter
\begin{equation}
    U \equiv \frac{\theta_S}{\theta_E} = \frac{R_S/d_S}{R_E/d_L},
\end{equation}
where $\theta_S$ is the angular size of the source. The finite size effects are most prominent in the regime where $U \gg 1$, but are not negligible even when $U < 1$. The magnification is given by integrating Eq. (\ref{Ageo}) over the source star in the plane of the lensing PBH
\begin{equation}
    A_{finite}(u,U) \equiv \frac{1}{\pi U^2} \int_{|\textbf{y}| \leq U} d^2\textbf{y}A_{geo}(|\textbf{u} - \textbf{y}|).
    \label{Afinite}
\end{equation}

\begin{figure}[!t]
    \centering
    \includegraphics[scale=0.28]{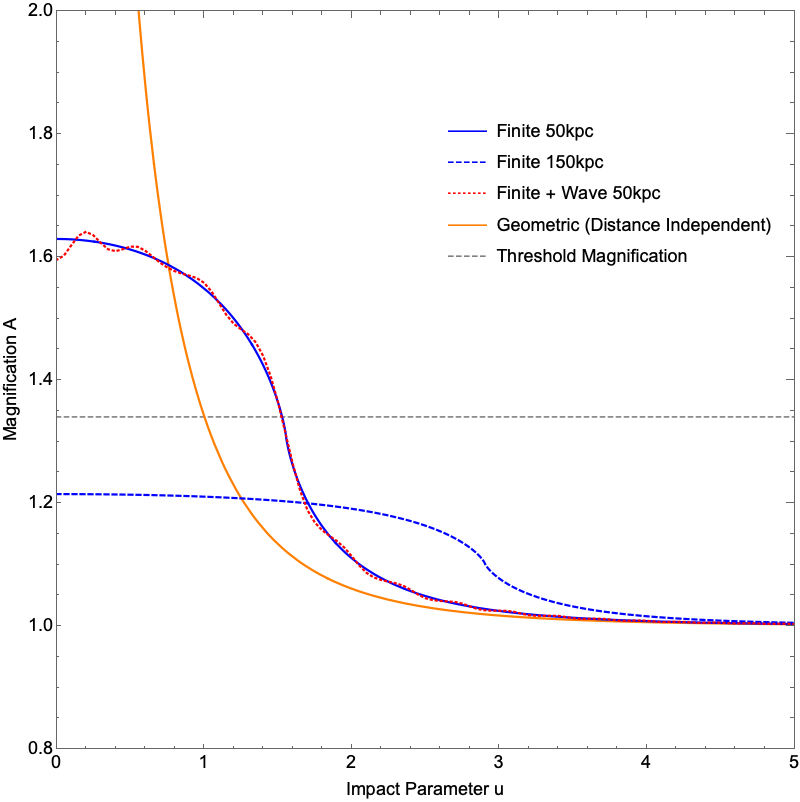}
    \caption{The threshold impact parameter values for geometric (orange), finite (blue), and finite + wave (red) models with a source star of radius $R_\odot$. In the geometric approximation, the threshold impact parameter is always 1, regardless of lens distance. When including finite size effects, the threshold impact parameter now depends on distance. If we move the lens further away, that is, closer to the source star, at a certain distance there will no longer be any detectable magnification as shown by the dashed blue line. The wave effects are small corrections to the finite size effects and are thus not considered for the final results of this paper.}
        \label{MagImpactComparison}
\end{figure}

Under the geometric approximation, Eq.~(\ref{Ageo}), the threshold magnification corresponds to a threshold impact parameter value of $u = 1$. When taking into account the finite size effects, the threshold impact parameter is different from unity in general. Following the prescription of \cite{sugiyama_revisiting_2019}, we calculate the value of the impact parameter that corresponds to a detectable event by setting $A_{finite} = 1.34$ for a particular set of parameters $d_L$, $M_{PBH}$, $r_{source}$, and solving for $u_{thresh}$ in Eq. (\ref{Afinite}). The dependence of $u_{thresh}$ on these parameters is shown in Fig.~\ref{MagImpactComparison}. 

% \begin{figure*}%
% \centering
% \subfigure[One]{%
% \label{MagImpactComparisonDouble}%
% \includegraphics[scale=0.24]{figures/MagImpactSquare.png}}%
% \qquad
% \subfigure[Two]{%
% \label{DistImpactDouble}%
% \includegraphics[scale=0.24]{figures/DistImpactSquare.pdf}}%
% \caption{Double Figure}
% \label{uthreshDouble}
% \end{figure*}

\section{Source Size Revisited\protect\\}
\label{sourcesize}

\begin{figure}[!t]
    \centering
    \includegraphics[scale=0.28]{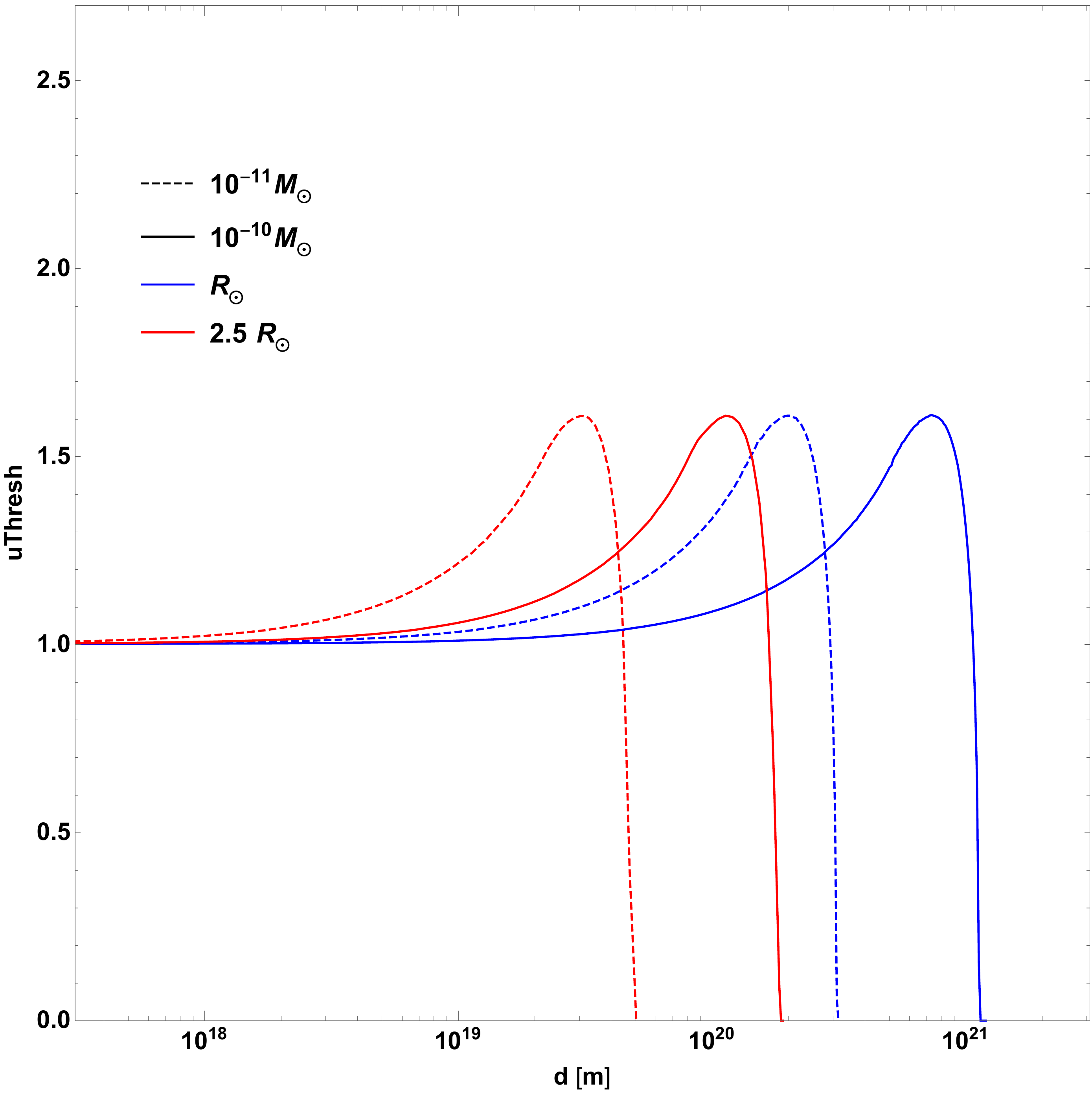}
    \caption{The dependence of the threshold impact parameter on $M_{PBH}$ and $r_{source}$ for 1 solar radius (blue), and 2.5 solar radii (red). The solid lines correspond to $M_{PBH} = 10^{-10} M_\odot$ while the dashed lines correspond to $M_{PBH} = 10^{-11} M_\odot$. For larger stars, the finite size effects become important at a smaller distance. Similarly, for lighter PBHs, the finite size effects are more dominant. Considering the population of stars in M31, this results in little to no detectable magnification for PBHs close to M31 unless they are well above asteroid mass.}
        \label{DistImpact}
\end{figure}

The HSC constraints assume a source size of $R=R_\odot$ for simplicity \cite{niikura_microlensing_2019}, but  observations are much more sensitive to significantly larger stars. To show this, we assume a black-body radiation spectrum which implies

\begin{equation}
    R = \sqrt{\frac{L_{bol}}{4\pi T^4 \sigma}},
    \label{Blackbody}
\end{equation}
where $\sigma$ is Wien's constant. We find that the apparent magnitude of the Sun would be $\approx$ 29 mag in the r-band were it located at the center of M31. Considering the detection efficiency of the HSC (30-20\% for $m_r = 25-26$ mag,  and 70-60\% for $m_r = 23-24$ mag) \cite{niikura_microlensing_2019}, a main-sequence star located in M31 would need to have a luminosity much greater than that of the Sun in order for a lensing event to be detectable. This implies that the HSC is more sensitive to larger main-sequence stars in general, according to Eq. (\ref{Blackbody}). This can be seen in Fig.~\ref{DistImpact} where for larger sources and/or smaller PBHs, there is no appreciable magnification, and therefore no valid $u_{thresh}$, unless the PBH is very far from the source.

If the source star were off-main sequence, the finite size effects would become even more significant. Therefore, here we aim to derive the corrected constraints from HSC observations using the population of stars in M31 that were both in the field of view and had light curves that could be successfully recovered by observations.

We use the catalog of stars from the Panchromatic Hubble Andromeda Treasury (PHAT) survey to find the population of stars in M31 which can have detectable microlensing events \cite{williams_panchromatic_2014, dalcanton_panchromatic_2012}. The PHAT survey resolved 117 million individual stars and partially overlaps with the HSC data in the disk region. However, the HSC is unable to resolve the fainter stars from the PHAT catalog. In order to ensure well-measured colors and coincidence between PHAT and microlensing-detectable stars, we use those stars that have a high signal-to-noise ratio and sufficient sharpness in each filter, as indicated by the GST tag, see \cite{williams_panchromatic_2014}.

The PHAT observations are grouped into 23 bricks. Each brick corresponds to a different region in M31, with higher numbered bricks corresponding to regions further from the galactic center. Since the HSC was unable to recover microlensing events in the bulge due to saturation of the detector, we remove this region from our analysis by considering only GST stars from bricks 7 and higher. We also perform a magnitude cut of $m \leq 26$ in the HST WFC F814W filter since this corresponds to the dimmest stars the HSC could observe in the r band. This preferentially eliminates small stars from our analysis which have lower luminosities on average. It should be noted that at many points in the HSC survey, the dimmest stars that could be observed were closer to $m = 24$. Thus the corrections made in this paper are on the conservative side in that the constraints would likely be weakened further by excluding those stars with magnitude $24 \leq m \leq 26$ which the HSC could not observe.

To determine the size of stars in the PHAT catalog, we use the MESA Isochrones and Stellar Tracks (MIST) stellar evolution package \cite{choi_mesa_2016, dotter_mesa_2016}. We generate isochrones for the non-rotating models in the HST ACS/WFC photometric system. We then compare each PHAT star to all the synthetic MIST stars using a nearest neighbors approach. This method works by finding those MIST stars that share sufficiently similar photomoetry to the PHAT star. A MIST star is considered sufficiently similar, or a neighbor, if it lies within $0.025$ in magnitude of a PHAT star using the apparent magnitudes of a star in each filter. This cutoff value was selected by calculating the number of PHAT stars with at least one nearest neighbor for various cutoff values. Above $\approx 0.025$, the number of stars with at least one neighbor doesn't increase significantly. This indicates that most of the stars that have a similar neighbor have already been found. Through this selection process, we found radius probability distributions for approximately 93\% of stars in the PHAT catalog. Of these, the standard deviation of the nearest neighbors was $0.4  R_\odot$ or less for 90\% of samples. This quality of fit is more than sufficient considering the small effect this uncertainty has on the final results (see Section \ref{Constraints} for details).

We use the bolometric luminosity and temperature of each synthetic MIST star to construct a probability distribution for the radius of each PHAT star. Because the MIST isochrones artificially contain a large number of high mass stars, we weight the synthetic stars using the Chabrier initial mass function, which disfavors very high mass stars \cite{chabrier_galactic_2003}. We also weight the data by the implied distance of the synthetic stars. By comparing the distance modulus of the synthetic stars to the real PHAT stars, we can compute how far away the synthetic star would be. By comparing this to the actual distance to an M31 star (following \cite{niikura_microlensing_2019}, the distance to each star is assumed to be the measured distance to the galactic center of M31, $\approx$ 770 kpc), we can determine how good a fit the neighboring synthetic stars are and weight them accordingly. Both of these weighting schemes are used for the remainder of this paper.

\begin{figure}
    \centering
    \includegraphics[scale=0.4]{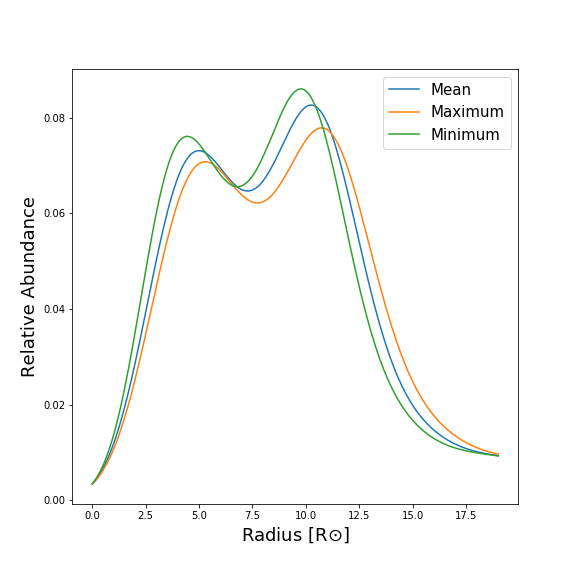}
    \caption{The population of stars in M31 which could have microlensing light curves resolvable by the HSC survey. Larger stars tend to have greater total luminosity in general and are therefore the easiest to detect.  The blue line shows the values used for our benchmark constraints. The uncertainty estimation, represented by the green and orange curves, comes from using the largest or smallest radius estimate for each star. See Section \ref{sourcesize} for details.}
    \label{kde}
\end{figure}

The result is the distribution of star sizes shown in the kernel density distribution of Fig.~\ref{kde}. The mean value of the neighboring stars is the distribution used to calculate the benchmark constraints. To see how much the constraints vary with this estimate, we also make a distribution using the largest and smallest radius stars within each set of neighbors (orange and green curves). 

The first peak around $4 R_\odot$ comes from main sequence stars, the most abundant branch. In order to  enable HSC-detectable lensing events, main sequence stars must be significantly more luminous than the Sun. Hence, according to Eq. (\ref{Blackbody}), the stars we consider are typically larger than the Sun, leading to the observed peak. The second peak around $10 R_\odot$ likely stems from the over-density of red giant branch and asymptotic giant branch stars in the disk of M31 \cite{dalcanton_panchromatic_2012, gordon_panchromatic_2016}. These stars are more luminous in general, and are therefore more likely to be observable to the HSC. As a result, even though the abundance of smaller stars is greater, it is in fact the larger stars that contribute most to the constraints. 

\section{Constraints on PBH as DM:\protect\\}
\label{Constraints}

Following the derivation of the original HSC constraints \cite{niikura_microlensing_2019}, which assume an isotropic Maxwell velocity distribution, monochromatic mass spectrum, and a NFW density profile for the DM \cite{navarro_universal_1997}, the differential event rate for microlensing of a single star by a PBH is given by 
\begin{equation}
    \label{difRate}
    \begin{split}
     \frac{d\Gamma_{PBH}}{d\hat{t}} = 2 \frac{\Omega_{PBH}}{\Omega_{DM}} \int_0^{d_s}dd_L \int_0^{U_T}du_{min} \\ \frac{1}{\sqrt{u_T^2 - u_{min}^2}} \frac{\rho_{DM}(d_L)}{M_{PBH}v_c^2(d_L)}v^4 \exp \Big[ -\frac{v^2}{v_c^2(d_L)}\Big],
    \end{split}
\end{equation}
where $v = 2R_E \sqrt{u_T^2 - u_{min}^2}/\hat{t}$ is the transverse velocity of the PBH, $d_L$ is the distance to the lensing PBH, and $d_S$ is the distance to the source star. Formally, a Maxwell velocity distribution is inconsistent with the assumption of a NFW density profile (see e.g. \cite{alcock_theory_1995}), but we assume this will have a negligible effect on the final constraints.

The duration of observation was 7 hours, of which the greatest sensitivity to detection of events was from $0.07$ hours to $3$ hours. We integrate over the observation time to find the total expected rate of events. To derive constraints, we assume microlensing events follow a Poisson distribution and compare the actual number of detections to the expected number of detections. The probability to observe $N_{obs}$ events is 

\begin{equation}
    P(k = N_{obs}|N_{exp}) = \frac{N_{exp}^k}{k!}e^{-N_{exp}}.
\end{equation} 
The HSC found one candidate event, so the 95\% confidence interval corresponds to $P(k=0) + P(k=1) \geq 0.05$, which leads to the condition $N_{exp} < 4.74$. 

\begin{figure}
    \centering
    \includegraphics[scale=0.28]{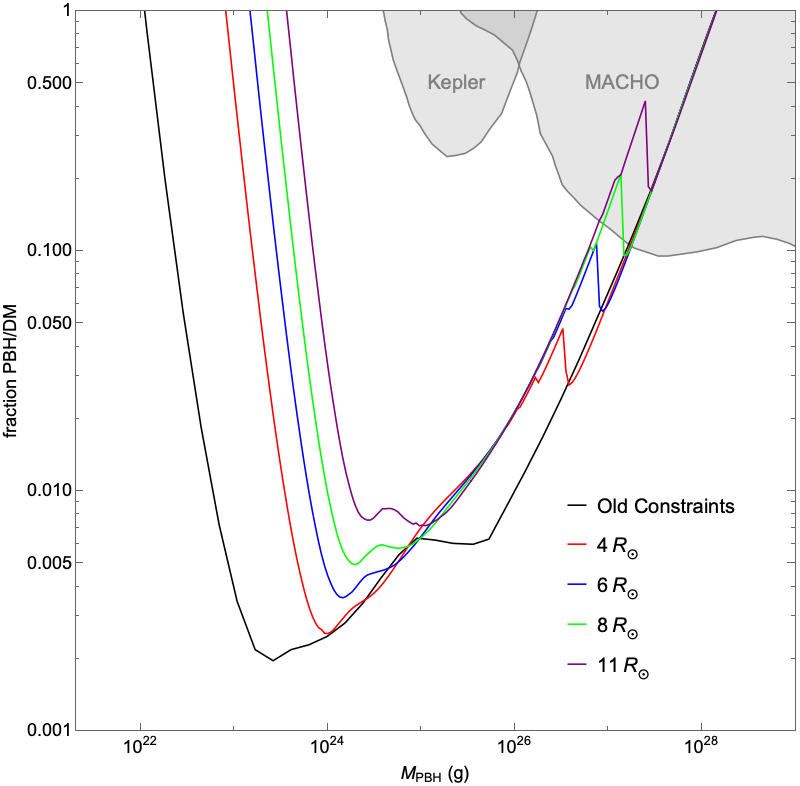}
    \caption{The constraints for example individual radii of a source star. The jagged feature arises from the point at which the DM density contribution from PBH in M31 in Eq. (\ref{difRate}) in the main text becomes negligible. Lower mass PBHs must be located close to the Milky Way in order to facilitate a detectable lensing event. The sharpness of this feature is averaged out in the benchmark constraints.}
    \label{CompareConstraints}
\end{figure}

The stellar population in Fig.~ \ref{kde} is sorted into linearly spaced bins up to $20 R_\odot$, at which point we use logarithmically spaced bins for the remaining few large, sparsely distributed stars. Constraints are generated by performing the integral in Eq. (\ref{difRate}) for each bin. Examples of how the constraints change with source size are shown in Fig.~\ref{CompareConstraints}. The benchmark constraints are then generated by appropriately weighting through a harmonic mean each of the constraints from a given stellar size according to the abundance of stars within the corresponding bin. 

\begin{figure*}
    \centering
    \includegraphics[scale=0.4]{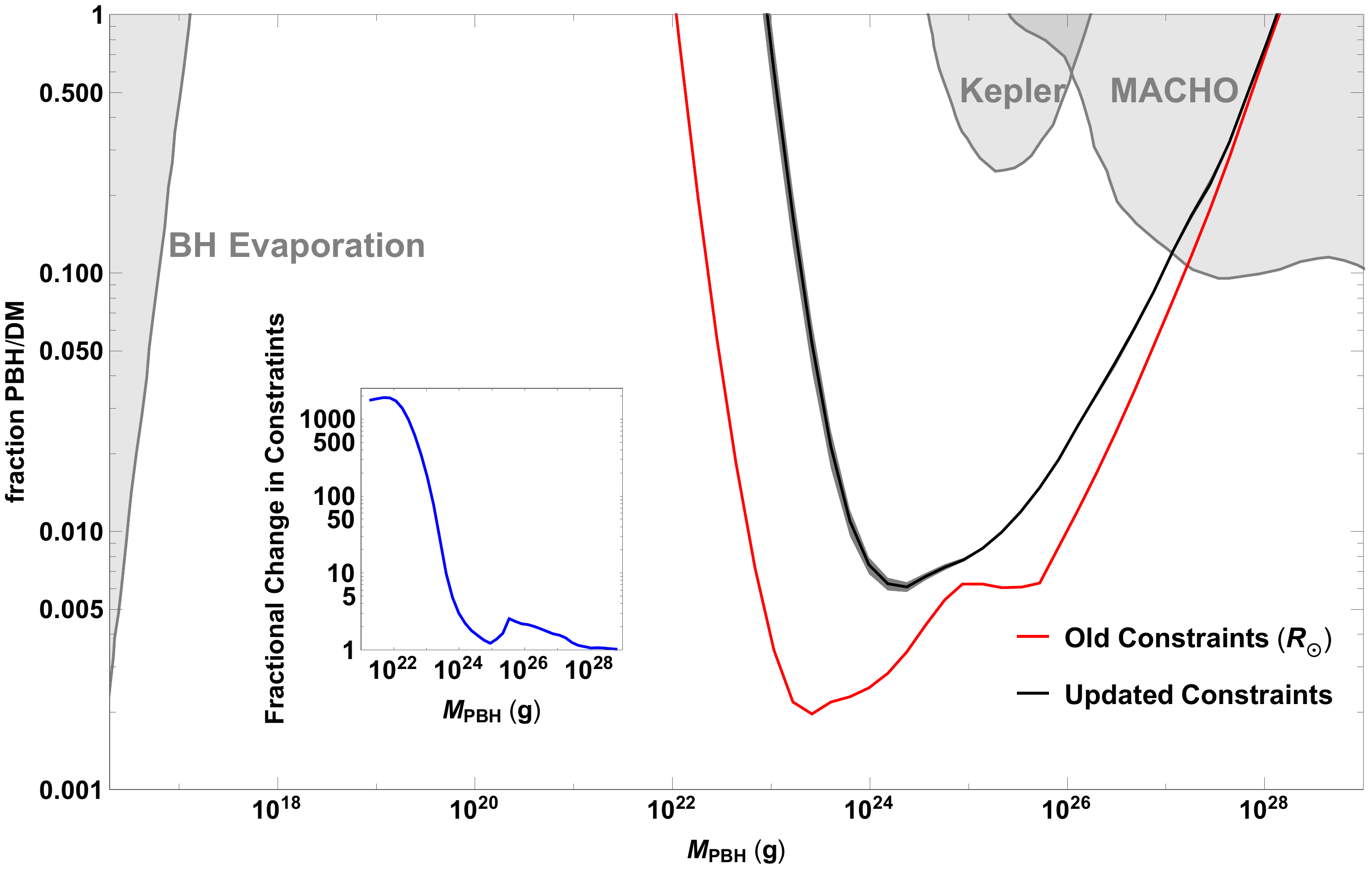}
    \caption{The constraints on primordial black holes as dark matter. The black line is the benchmark constraint and the primary result of this paper. The gray shading comes from the uncertainty in determining the stellar size distribution. The red line is the previous HSC constraint which includes finite size effects, but assumes that all stars in M31 have a radius of $R_\odot$ \cite{niikura_microlensing_2019}. The other constraints come from extra-galactic gamma ray searches from evaporating black holes \cite{carr_new_2010} as well as microlensing searches using Kepler \cite{griest_new_2013}, and MACHO/EROS \cite{tisserand_limits_2007}.} 
    \label{BenchmarkConstraints}
\end{figure*}

Fig.~\ref{BenchmarkConstraints} shows our key results on the plane of the fraction of PBH to the total DM abundance versus the PBH mass. The inset shows the fractional change in the constraints compared to the original ones in Ref~\cite{niikura_microlensing_2019}. After taking into proper consideration the distribution of stellar sizes, the HSC constraints are up to three orders of magnitude weaker than in the original estimate with stellar sizes all set to one solar radius. As a result, the range of PBH masses where PBHs can be the DM increases by almost one order or magnitude.

The correction to the constraints due to larger actual stellar sizes is significant, especially in the low mass range. To appreciate this, we point out the scaling of the constraints with source radius (see also Ref.~\cite{montero-camacho_revisiting_2019}): if the source radius doubles, in order to keep the same parameter $U$, the lens would need to be brought closer to Earth according to:
\begin{equation}
    U = \frac{\theta_S}{\theta_E} = \frac{R_S d_L}{R_E d_S} = R_S \Big(\frac{x}{1-x}\Big)^{1/2},
\end{equation}
where $x = d_L/d_S$. If x is small, which is a good approximation for low mass PBHs since $u_T$ is 0 unless $d_L \ll d_S$, this corresponds to the lens being brought a factor of 4 closer to the Earth. This, in turn, means that the Einstein radius of the PBH will halve. Since $v \propto R_E$, our rate of observed lensing events will go down by a factor of $1/2^6 = 1/64$ for low mass PBHs \cite{montero-camacho_revisiting_2019}. This explains why in the low mass regime, the finite size effects can change by such a large margin for a relatively small change in $R_S$.

The finite size effects also weaken the constraints for intermediate mass black holes. The finite size effects are greatest when the lensing PBH is close to the source star. That is, as $d_L$ approaches $d_S$, the finite size effect washes out magnification for all masses of PBH. This is also the region where the DM density contribution from M31 in Eq. (\ref{difRate}) is the greatest, leading to a noticeable weakening of the constraints for all but the high mass region of the relevant parameter space.

The uncertainty in the constraints given the uncertainty in stellar sizes we find is shown in the gray shading in Fig.~\ref{BenchmarkConstraints}, obtained using the smallest and largest nearest neighbor given by the MIST comparison. In the high mass region, the uncertainty in the constraint is minuscule, since finite size effects are less important in this region. At lower masses, the uncertainty is noticeable, but still small. When using the minimum estimate for the size of a source star, any given bin will lose some number of stars to a smaller bin, but will also gain a number of stars from a larger bin. The net effect is slight, but reassuring as it suggests that the number and size of bins matters much less than the overall distribution of stars. 

\section{Conclusions}

Our study illustrates how careful consideration of finite-size source effects leads to the swath of parameter space where PBH can be the totality of the DM being significantly larger than previously thought. This, in turn, begs for further studies as to how to best explore the region of sub-planetary mass PBHs. We incidentally note that our results apply to any sufficiently compact DM candidate in this mass range, including e.g. dark quark nuggets, strangelets, axion miniclusters, or axion stars.

In general, optimal observing strategies would include long observations of the greatest possible number of stars, with a high enough cadence to enable detection of the shortest duration lensing events. Using a shorter wavelength filter to mitigate wave effects would also expand the testable PBH mass region, as suggested by \cite{sugiyama_revisiting_2019}. 

This work emphasizes how the observed stars need to be luminous enough in the band of observation to obtain a sufficient expected number of events. Additionally, isolating sources with smaller radii and observing very far away sources would limit the finite-size effects and improve constraints. But this proves challenging when trying to isolate individual stars and achieve the best possible image resolution. Indeed, finding the ideal candidates for observation is a critical step to improving microlensing constraints and is a subject of our future work. 

\vspace*{.5cm}

 \begin{acknowledgments}
NS and SP are partly supported by the U.S. Department of Energy grant number de-sc0010107.
 \end{acknowledgments}

%\nocite{*}

\bibliographystyle{unsrt}
\bibliography{references}% Produces the bibliography via BibTeX.

\begin{thebibliography}{10}

\bibitem{arcadi_waning_2017}
Giorgio Arcadi, Maíra Dutra, Pradipta Ghosh, Manfred Lindner, Yann Mambrini,
  Mathias Pierre, Stefano Profumo, and Farinaldo~S. Queiroz.
\newblock The {Waning} of the {WIMP}? {A} {Review} of {Models}, {Searches}, and
  {Constraints}.
\newblock March 2017.

\bibitem{carr_new_2010}
B.~J. Carr, Kazunori Kohri, Yuuiti Sendouda, and Jun'ichi Yokoyama.
\newblock New cosmological constraints on primordial black holes.
\newblock {\em Phys.Rev.}, D81:104019, 2010.

\bibitem{boudaud_voyager_2019}
Mathieu Boudaud and Marco Cirelli.
\newblock Voyager 1
  \$\{e\}{\textasciicircum}\{{\textbackslash}ifmmode{\textbackslash}pm{\textbackslash}else{\textbackslash}textpm{\textbackslash}fi\{\}\}\$
  {Further} {Constrain} {Primordial} {Black} {Holes} as {Dark} {Matter}.
\newblock {\em Physical Review Letters}, 122(4):041104, January 2019.

\bibitem{bird_did_2016}
Simeon Bird, Ilias Cholis, Julian~B. Muñoz, Yacine Ali-Haïmoud, Marc
  Kamionkowski, Ely~D. Kovetz, Alvise Raccanelli, and Adam~G. Riess.
\newblock Did {LIGO} detect dark matter?
\newblock March 2016.

\bibitem{poulin_cmb_2017}
Vivian Poulin, Pasquale~D. Serpico, Francesca Calore, Sebastien Clesse, and
  Kazunori Kohri.
\newblock {CMB} bounds on disk-accreting massive {Primordial} {Black} {Holes}.
\newblock {\em Physical Review D}, 96(8):083524, October 2017.
\newblock arXiv: 1707.04206.

\bibitem{inoue_new_2017}
Yoshiyuki Inoue and Alexander Kusenko.
\newblock New {X}-ray bound on density of primordial black holes.
\newblock {\em Journal of Cosmology and Astroparticle Physics},
  2017(10):034--034, October 2017.
\newblock arXiv: 1705.00791.

\bibitem{green_microlensing_2016}
Anne~M. Green.
\newblock Microlensing and dynamical constraints on primordial black hole dark
  matter with an extended mass function.
\newblock {\em Physical Review D}, 94(6):063530, September 2016.
\newblock arXiv: 1609.01143.

\bibitem{ulmer_femtolensing:_1995}
A.~Ulmer and J.~Goodman.
\newblock Femtolensing: {Beyond} the {Semi}-{Classical} {Approximation}.
\newblock {\em The Astrophysical Journal}, 442:67, March 1995.
\newblock arXiv: astro-ph/9406042.

\bibitem{katz_femtolensing_2018}
Andrey Katz, Joachim Kopp, Sergey Sibiryakov, and Wei Xue.
\newblock Femtolensing by {Dark} {Matter} {Revisited}.
\newblock {\em Journal of Cosmology and Astroparticle Physics},
  2018(12):005--005, December 2018.
\newblock arXiv: 1807.11495.

\bibitem{niikura_microlensing_2019}
Hiroko Niikura, Masahiro Takada, Naoki Yasuda, Robert~H. Lupton, Takahiro Sumi,
  Surhud More, Toshiki Kurita, Sunao Sugiyama, Anupreeta More, Masamune Oguri,
  and Masashi Chiba.
\newblock Microlensing constraints on primordial black holes with the
  {Subaru}/{HSC} {Andromeda} observation.
\newblock {\em Nature Astronomy}, 3(6):524--534, June 2019.
\newblock arXiv: 1701.02151.

\bibitem{sugiyama_revisiting_2019}
Sunao Sugiyama, Toshiki Kurita, and Masahiro Takada.
\newblock Revisiting the wave optics effect on primordial black hole
  constraints from optical microlensing search.
\newblock {\em arXiv:1905.06066 [astro-ph]}, May 2019.
\newblock arXiv: 1905.06066.

\bibitem{montero-camacho_revisiting_2019}
Paulo Montero-Camacho, Xiao Fang, Gabriel Vasquez, Makana Silva, and
  Christopher~M. Hirata.
\newblock Revisiting constraints on asteroid-mass primordial black holes as
  dark matter candidates.
\newblock {\em arXiv:1906.05950 [astro-ph]}, June 2019.
\newblock arXiv: 1906.05950.

\bibitem{paczynski_gravitational_1986}
B.~Paczynski.
\newblock Gravitational {Microlensing} at {Large} {Optical} {Depth}.
\newblock {\em ApJ}, 301:503, February 1986.

\bibitem{nakamura_wave_1999}
Takahiro~T. Nakamura and Shuji Deguchi.
\newblock Wave {Optics} in {Gravitational} {Lensing}.
\newblock {\em Progress of Theoretical Physics Supplement}, 133:137--153,
  January 1999.

\bibitem{takahashi_wave_2003}
Ryuichi Takahashi and Takashi Nakamura.
\newblock Wave {Effects} in the {Gravitational} {Lensing} of {Gravitational}
  {Waves} from {Chirping} {Binaries}.
\newblock {\em The Astrophysical Journal}, 595(2):1039--1051, October 2003.

\bibitem{matsunaga_finite_2006}
Norihito Matsunaga and Kazuhiro Yamamoto.
\newblock The finite source size effect and wave optics in gravitational
  lensing.
\newblock {\em Journal of Cosmology and Astroparticle Physics},
  2006(01):023--023, January 2006.

\bibitem{witt_can_1994}
Hans~J. Witt and Shude Mao.
\newblock Can lensed stars be regarded as pointlike for microlensing by
  {MACHOs}?
\newblock {\em The Astrophysical Journal}, 430:505--510, August 1994.

\bibitem{williams_panchromatic_2014}
Benjamin~F. Williams, Dustin Lang, Julianne~J. Dalcanton, Andrew~E. Dolphin,
  Daniel~R. Weisz, Eric~F. Bell, Luciana Bianchi, Nell Byler, Karoline~M.
  Gilbert, Léo Girardi, Karl Gordon, Dylan Gregersen, L.~C. Johnson, Jason
  Kalirai, Tod~R. Lauer, Antonela Monachesi, Philip Rosenfield, Anil Seth, and
  Eva Skillman.
\newblock {THE} {PANCHROMATIC} {HUBBLE} {ANDROMEDA} {TREASURY}. {X}.
  {ULTRAVIOLET} {TO} {INFRARED} {PHOTOMETRY} {OF} 117 {MILLION} {EQUIDISTANT}
  {STARS}.
\newblock {\em The Astrophysical Journal Supplement Series}, 215(1):9, October
  2014.

\bibitem{dalcanton_panchromatic_2012}
Julianne~J. Dalcanton, Benjamin~F. Williams, Dustin Lang, Tod~R. Lauer,
  Jason~S. Kalirai, Anil~C. Seth, Andrew Dolphin, Philip Rosenfield, Daniel~R.
  Weisz, Eric~F. Bell, Luciana~C. Bianchi, Martha~L. Boyer, Nelson Caldwell,
  Hui Dong, Claire~E. Dorman, Karoline~M. Gilbert, Léo Girardi, Stephanie~M.
  Gogarten, Karl~D. Gordon, Puragra Guhathakurta, Paul~W. Hodge, Jon~A.
  Holtzman, L.~Clifton Johnson, Søren~S. Larsen, Alexia Lewis, Jason~L.
  Melbourne, Knut A.~G. Olsen, Hans-Walter Rix, Keith Rosema, Abhijit Saha, Ata
  Sarajedini, Evan~D. Skillman, and Krzysztof~Z. Stanek.
\newblock {THE} {PANCHROMATIC} {HUBBLE} {ANDROMEDA} {TREASURY}.
\newblock {\em The Astrophysical Journal Supplement Series}, 200(2):18, May
  2012.

\bibitem{choi_mesa_2016}
Jieun Choi, Aaron Dotter, Charlie Conroy, Matteo Cantiello, Bill Paxton, and
  Benjamin~D. Johnson.
\newblock Mesa {Isochrones} and {Stellar} {Tracks} ({MIST}). {I}.
  {Solar}-scaled {Models}.
\newblock {\em The Astrophysical Journal}, 823:102, June 2016.

\bibitem{dotter_mesa_2016}
Aaron Dotter.
\newblock {MESA} {Isochrones} and {Stellar} {Tracks} ({MIST}) 0: {Methods} for
  the {Construction} of {Stellar} {Isochrones}.
\newblock {\em The Astrophysical Journal Supplement Series}, 222:8, January
  2016.

\bibitem{chabrier_galactic_2003}
Gilles Chabrier.
\newblock Galactic {Stellar} and {Substellar} {Initial} {Mass} {Function}.
\newblock {\em Publications of the Astronomical Society of the Pacific},
  115:763--795, July 2003.

\bibitem{gordon_panchromatic_2016}
Karl~D. Gordon, Morgan Fouesneau, Heddy Arab, Kirill Tchernyshyov, Daniel~R.
  Weisz, Julianne~J. Dalcanton, Benjamin~F. Williams, Eric~F. Bell, Luciana
  Bianchi, Martha Boyer, Yumi Choi, Andrew Dolphin, Leo Girardi, David~W. Hogg,
  Jason~S. Kalirai, Maria Kapala, Alexia~R. Lewis, Hans-Walter Rix, Karin
  Sandstrom, and Evan~D. Skillman.
\newblock The {Panchromatic} {Hubble} {Andromeda} {Treasury} {XV}. {The}
  {BEAST}: {Bayesian} {Extinction} and {Stellar} {Tool}.
\newblock {\em The Astrophysical Journal}, 826(2):104, July 2016.
\newblock arXiv: 1606.06182.

\bibitem{navarro_universal_1997}
Julio~F. Navarro, Carlos~S. Frenk, and Simon D.~M. White.
\newblock A {Universal} {Density} {Profile} from {Hierarchical} {Clustering}.
\newblock {\em ApJ}, 490(2):493--508, December 1997.

\bibitem{alcock_theory_1995}
C.~Alcock, R.~A. Allsman, T.~S. Axelrod, D.~P. Bennett, K.~H. Cook, N.~W.
  Evans, K.~C. Freeman, K.~Griest, J.~Jijina, M.~Lehner, S.~L. Marshall,
  S.~Perlmutter, B.~A. Peterson, M.~R. Pratt, P.~J. Quinn, A.~W. Rodgers, C.~W.
  Stubbs, W.~Sutherland, and {MACHO Collaboration}.
\newblock Theory of {Exploring} the {Dark} {Halo} with {Microlensing}. {I}.
  {Power}-{Law} {Models}.
\newblock {\em The Astrophysical Journal}, 449:28, August 1995.

\bibitem{griest_new_2013}
Kim Griest, Agnieszka~M. Cieplak, and Matthew~J. Lehner.
\newblock New {Limits} on {Primordial} {Black} {Hole} {Dark} {Matter} from an
  {Analysis} of {Kepler} {Source} {Microlensing} {Data}.
\newblock {\em Physical Review Letters}, 111(18):181302, October 2013.

\bibitem{tisserand_limits_2007}
P.~Tisserand, L.~Le Guillou, C.~Afonso, J.~N. Albert, J.~Andersen, R.~Ansari,
  E.~Aubourg, P.~Bareyre, J.~P. Beaulieu, X.~Charlot, C.~Coutures, R.~Ferlet,
  P.~Fouqué, J.~F. Glicenstein, B.~Goldman, A.~Gould, D.~Graff, M.~Gros,
  J.~Haissinski, C.~Hamadache, J.~de~Kat, T.~Lasserre, E.~Lesquoy, C.~Loup,
  C.~Magneville, J.~B. Marquette, E.~Maurice, A.~Maury, A.~Milsztajn,
  M.~Moniez, N.~Palanque-Delabrouille, O.~Perdereau, Y.~R. Rahal, J.~Rich,
  M.~Spiro, A.~Vidal-Madjar, L.~Vigroux, and S.~Zylberajch.
\newblock Limits on the {Macho} {Content} of the {Galactic} {Halo} from the
  {EROS}-2 {Survey} of the {Magellanic} {Clouds}.
\newblock {\em Astronomy \& Astrophysics}, 469(2):387--404, July 2007.
\newblock arXiv: astro-ph/0607207.

\end{thebibliography}

% \vspace*{3cm}
% \onecolumngrid
% \section*{Supplemental Material for 'Updated Constraints on Asteroid-Mass Primordial Black Holes as Dark Matter'}% Force line breaks with \\

% %\clearpage

% \begin{figure}
%     \centering
%     \includegraphics[scale=0.4]{figures/BestVsWorstCutHist.png}
%     \caption{Binned distribution corresponding to the kernel density of Fig.~\ref{kde} in the main text for the size of stars contributing to the HSC constraints on PBH.}
%     \label{BestVsWorst}
% \end{figure}

%see last figure of http://spiff.rit.edu/classes/phys440/lectures/size/size.html for comparison

\end{document}